# Bose-Einstein Condensation of Nonequilibrium Magnons in Confined Systems

Morteza Mohseni,[1,*] Alireza Qaiumzadeh,[2] Alexander A. Serga,[1] Arne Brataas,[2] Burkard Hillebrands[1] and Philipp Pirro[1]

[1] Fachbereich Physik and Landesforschungszentrum OPTIMAS, Technische Universität Kaiserslautern,
67663 Kaiserslautern, Germany
[2] Center for Quantum Spintronics, Department of Physics, Norwegian University of Science and Technology,
NO-7491 Trondheim, Norway

We study the formation of a room temperature magnon Bose-Einstein condensate (BEC) in nanoscopic systems and demonstrate that its lifetime is influenced by the spatial confinement. We predict how dipolar interactions and nonlinear magnon scattering assist in the generation of a metastable magnon BEC in energy-quantized nanoscopic devices. We verify our prediction by a full numerical simulation of the Landau-Lifshitz-Gilbert equation and demonstrate the generation of magnon BEC in confined insulating magnets of yttrium iron garnet. We directly map out the nonlinear magnon scattering processes behind this phase transition to show how fast quantized thermalization channels allow the BEC formation in confined structures. Based on our results, we discuss a new mechanism to manipulate the BEC lifetime in nanoscaled systems. Our study greatly extends the freedom to study dynamics of magnon BEC in realistic systems and to design integrated circuits for BEC-based applications at room temperature.

Bose-Einstein condensates (BEC) are exotic quantum states of matter [1-2]. The associated phase transition manifests as an abrupt macroscopic growth in the population of particles in the quantum ground state. Similar phenomena but with different condensation mechanisms occur for bosonic quasiparticles (QPs) which are elementary excitations of a solid-state system, such as photons, excitons, polaritons, and magnons [3-9]. Magnons, the quanta of spin-waves (SWs) carrying a quantized amount of energy and momentum, are the low energy spin excitations of a magnetically ordered system [10]. The BEC of magnons can be achieved by e.g., tuning the effective chemical potential of weakly interacting magnons via increasing their density at room temprature [7-8].

Starting with the first observation of nonequilibrium magnon BEC in macroscopic Yttrium Iron Garnet (YIG) films at room temperature, almost all experimental studies consider similar system sizes, see e.g., Refs. [6-8, 11] Indeed, YIG films are the most suitable hosts to explore similar phenomena due to their ultra-low magnon relaxation rates [12]. However, the use of relatively thick films does not allow the investigation of this phenomenon in quantized states since the magnon bands in these systems are quasi-continuous and show crossings and hybridization [12-15]. This limitation prevents an explicit understanding of the nonlinear scattering processes (e.g. among modes) that leads to the formation of a magnon BEC. Moreover, the spectral limitations of the experimental techniques hinder a full understanding of the details of the nonlinear scattering processes behind this phenomenon. Furthermore, the presented analytical models use many approximations since the inclusion of all modes and nonlinear scattering is challenging [8-9, 13-14, 16-17]. This leads to the lack of access to study this phenomenon thoroughly without further assumptions, suggesting advanced numerical analysis is required for a more precise treatment of such dynamics.

To overcome these challenges, downscaling macroscopic systems toward nanometer sizes to enhance the quantization of the magnon energy levels is necessary [18-19]. Downscaling is also very important for applications such as data transport via spin superfluidity [20-26]. However, so far only a few studies have been conducted to explore the possible formation of magnon BEC in nanoscopic systems. In addition, the impact of the spatial confinement on the dynamics of magnon gases and BEC is not well understood. Indeed, this is an essential ingredient in device design to harness unique properties of magnon BEC in systems on a chip and integrated circuits. Besides, the recent discovery of a magnon BEC in a nanoscopic YIG thin film structure and nonlinear SW dynamics in such systems further motivates us to better understand the nonlinear scattering processes behind such a phase transition in similar devices [27,28].

Here, we address these questions by studying the dynamics of magnon gases and BEC in nanoscopic YIG devices. We show that the confinement enhances the stability and increases the lifetime of the nonequilibrium magnon BEC in ultrathin YIG structures. Indeed, such a metastable state is influenced significantly by the dipolar interactions. Therefore, since the effects of dipolar interactions can be more pronounced if the lateral dimensions of a YIG thin film structure are confined, we predict that a magnon BEC can form for a longer time.

We verify our prediction by solving the Landau–Lifshitz equation including Gilbert damping and thermal excitations in the geometry of a nanoscopic YIG conduit without any approximations. We illustrate how using this method enables an intuitive understanding of the dynamics behind such nonlinear processes. It will be shown that the injection of a sufficiently high density of magnons using parametric pumping leads to a sequence of nonlinear scattering events that redistribute the energy among magnon states and form a magnon condensate. We map out the sequences of the scattering processes and show that the thermalization time of the magnons in confined systems with dilute spectra is much smaller than in macroscopic samples [11].

We consider a magnetic thin film strcuture with periodic boundary conditions in the $x$-direction, a finite thickness $d$ in the $z$-direction, and a finite width $w$ in the $y$-direction, as shown in the inset of Fig. 1. If the system is magnetized along the $x$-direction and magnons propagate parallel to the magnetization direction, the lowest magnon energy band ($n=0$) reads [18, 29],

$$\omega_{n=0}(k_x) = \sqrt{[\omega_H + \omega_M(\lambda^2 K^2 + F_{k_x}^{yy})][\omega_H + \omega_M(\lambda^2 K^2 + F_{k_x}^{zz})]} \quad (1)$$

where $k_x$ is the longitudinal component of the wavevector along the magnetization vector **M**, $\omega_H = \gamma B$, $\omega_M = \gamma \mu_0 M_s$, $\gamma$ is the gyromagnetic ratio, $B$ is the static applied magnetic field along the $x$ direction, $\mu_0$ is the permeability of free space, $K = \sqrt{k_x^2 + \kappa_y^2}$ with $\kappa_y = \pi/w$, and $\lambda = \sqrt{2A_{\text{exch}}/(\mu_0 M_s^2)}$ is the exchange length, with an exchange constant $A_{\text{exch}}$ and a saturation magnetization $M_s$. Here, $F_{k_x}^{yy}$ and $F_{k_x}^{zz}$ are the $y$ and $z$ components of the demagnetization tensors respectively [18]. In this Letter, we consider a YIG film with a fixed thickness of $d = 85$ nm, $A_{\text{exch}} = 3.5$ pJ/m, $M_s = 140$ kA/m and a Gilbert damping of $\alpha = 0.0002$ [18-19].

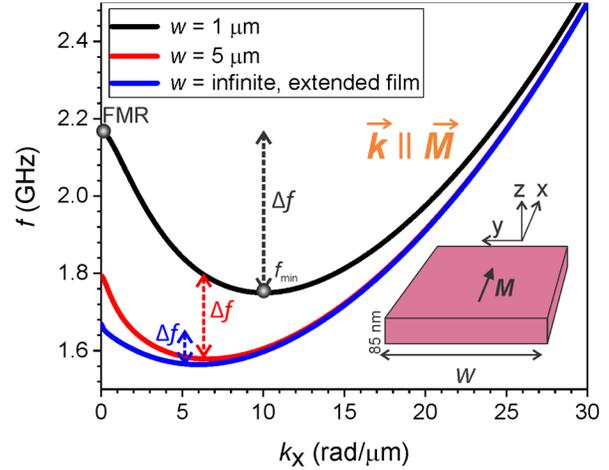

FIG. 1. Dispersion relation of the lowest energy magnon band of a YIG thin film structure with a thickness of $d = 85$ nm and different degrees of confinement in one lateral direction.

Figure 1 compares the lowest magnon mode of an infinitely extended film (blue curve) with the dispersion relations of two laterally confined conduits with a width of $w = 5$ μm (red curve) and $w = 1$ μm (black curve), all in the presence of an 18 mT bias field. The interplay between the dipolar and exchange interactions leads to the appearance of an energy minimum in the frequency spectrum denoted as $f_{\min}$, at $k_x = \pm Q$, the band bottom which corresponds to the quantum ground state of the system.

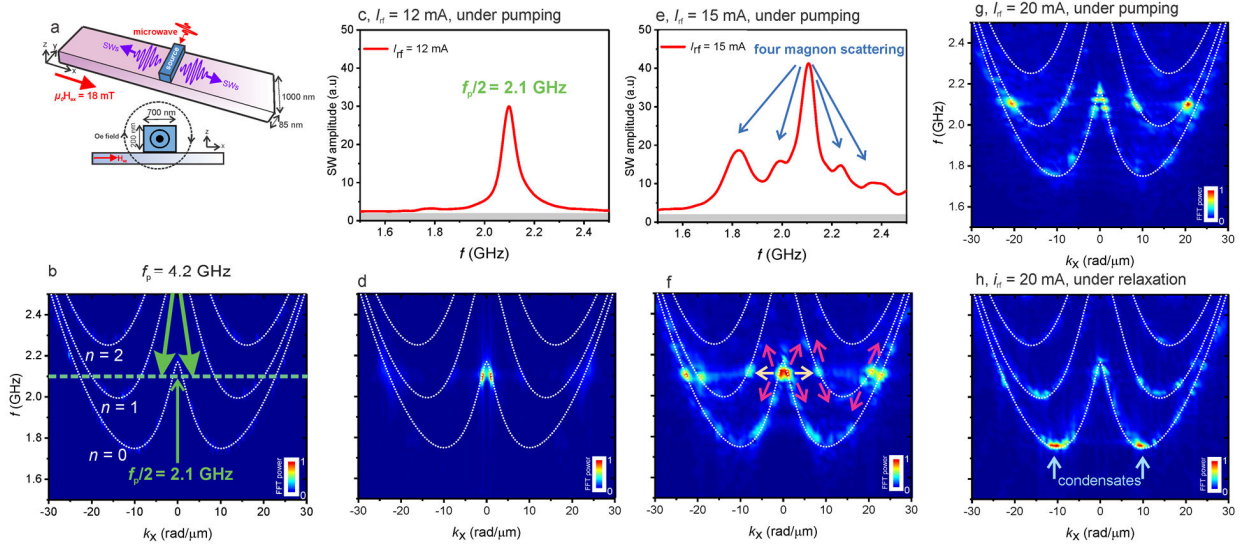

FIG. 2. (a) Schematic view of the sample under study. (b) Magnon band structure near the band bottom obtained from numerical simulations (color plot) and analytical calculations (dotted lines). The first three width modes ($n$ = 0, 1, 2) are present in the band structure. (c-d), (e-f): Frequency spectra and the magnon band structures during pumping when microwave currents of $I_{rf}$ = 12 mA and $I_{rf}$ = 15 mA are injected into the stripline, respectively. Thermal magnons are distinguished via gray color at (c-e). (g-h): Magnon band structure during pumping and relaxation, when the microwave current is $I_{rf}$ = 20 mA. Magnon condensation at the band bottom is visible during relaxation.

More interestingly, the lateral confinement leads to an enhanced dipolar interaction along the propagation direction and consequently a deeper band depth. In principle, lateral confinement leads to quantization of the SW modes across the width of the system. In longitudinally magnetized confined systems, spins at the edges of the conduits are (partially) pinned due to the dipolar interactions, or in other words, the demagnetization field. This means that the amplitude of the SW modes at the edges of the conduits is reduced. Considering the boundary conditions at the lateral edges of the conduit, and as can be seen from Eq. 1, the magnon band structure strongly depends on the $y$ and $z$ components of the demagnetization tensors $F_{k_x}^{yy}$ and $F_{k_x}^{zz}$. This means that the magnon band structure depends on the sample geometry since it determines the demagnetization tensors. The lateral component of the demagnetization tensor is close to zero in ultrathin and wide conduits e.g. $w$ = 10 μm in our study, and strongly increases with decreasing the width of the conduits due to a stronger quantization [18, 29]. With decreasing width, the contributions of the dynamic demagnetization fields arising from the edges of the waveguides are getting more important compared to the demagnetization fields from the surface. The larger demagnetization tensor in narrower conduits increases the frequency of the Ferromagnetic resonance (FMR) frequency at $k_x$ = 0 and consequently, shifts the frequency of the dipolar waves with smaller wave vectors to higher frequencies. This finally deepens the band depth. Note that for large wave numbers, the dipolar contributions to the magnon band structure are saturated and the exchange energy determines the frequency of the SWs which is not affected by the confinement. Thus, the band structure of conduits with different widths converges at high wave numbers. We define the band depth as the difference between the band bottom of the lowest mode and the ferromagnetic resonance (FMR) frequency as $\Delta f$ = $f_{FMR}$ - $f_{min}$ shown by the dashed arrows in Fig. 1.

As reflected by the observed finite lifetime of the magnon BEC [8, 27], the magnon BEC is a metastable state with respect to thermal activations or decay to the excited states [30-31]. This causes lower and upper limits of the magnon density required to form and destroy a magnon BEC, respectively. Indeed, one of the most prominent features of the metastability of the magnon BEC is its finite lifetime, even in the absence of Gilbert dissipation to the lattice. Therefore, we expect that its lifetime is enhanced in systems with a deeper band depth. This is caused by the fact that in such a system, there is a large energy separation between the condensate state at the band bottom and the excited states (e.g. the externally injected magnons or magnons scattered to higher energies than the quantum ground state) that cannot be exceeded by the magnon-magnon scattering within the weak interaction regime once the BEC is established [7-8, 30,31]. This energy gap between the condensed magnons and pumped magnons is larger if the band depth is larger and consequently, such a dissipation channel for the condensed magnons is weakened. Thus, the magnon BEC dissipates with a slower rate and possess a longer lifetime for larger band depth. In other words, this deeper band depth creates a larger energy barrier to forbid or at least to minimize the interactions of the magnons in the BEC with the injected magnons at higher energies close to the ferromagnetic resonance.

In our case, this is equivalent to systems with narrower lateral dimensions, i.e., a system with enhanced dipolar interactions along the propagation direction. Therefore, we expect that under this condition, the magnon condensate forms for a finite time, and its lifetime is influenced by the band depth. In the following, we verify this hypothesis.

We use MuMax 3.0 open source GPU-based software to numerically solve the LLG equation with thermal noise [32-33]. The system is a YIG conduit with a thickness of $d = 85$ nm and a lateral width of $w = 1$ μm as shown in Fig. 2a. A microwave pumping pulse with a duration of $t_{pump} = 50$ ns with a carrier frequency of $f_p = 4.2$ GHz is applied to the simulated microwave stripline (placed on top of the conduit) whose dynamic Oersted fields leads to parametric generation of the magnons at $f_p/2 = 2.1$ GHz that is slightly below the FMR frequency [34, 35-36]. This process has been studied experimentally and numerically in Ref. [35]. After the pumping pulse, we let the system relax for $t_{relax} = 50$ ns. We analyze the data during the pumping and relaxation periods separately. More details about the simulations can be found in [33, 36]. We first set the microwave current to $I_{rf} = 12$ mA, which produces a pumping field with an amplitude of $\mu_0 h_p = 8.9$ mT that is well above the threshold of the parametric generation of magnons in the system [36-37]. In Fig. 2c, the frequency spectrum directly shows the parametrically generated magnons at $f_p/2$. The magnon band structure under this pumping field as displayed in Fig. 2d shows that these magnons predominantly populate the dipolar branch of the first mode, due to its large ellipticity and highest coupling to the pumping field [36-37].

Next, we increase the microwave current to $I_{rf} = 15$ mA ($\mu_0 h_p = 11.2$ mT) in order to generate a higher density of magnons. Indeed, if the induced magnon density is high enough, nonlinear four-magnon scattering becomes pronounced [6-8]. Such nonlinear scattering processes, indicated by blue arrows in Fig. 2e, obey energy and momentum conservation laws [7-8, 34, 38],

$$\sum_{i=1}^{2} \hbar\omega_i = \sum_{j=1}^{2} \hbar\omega_j,$$
$$\sum_{i=1}^{2} \hbar\boldsymbol{k}_i = \sum_{j=1}^{2} \hbar\boldsymbol{k}_j \qquad (2)$$

where two incoming magnons with frequencies $\omega_i$ and momenta $\hbar\boldsymbol{k}_i$ scatter into two outgoing magnons with frequencies $\omega_j$ and momenta $\hbar\boldsymbol{k}_j$. However, the total number of magnons during this process is constant. We can distinguish two types of four-magnon scattering processes in the system as presented in Fig. 2f: *i*) four-magnon scattering processes in which two incoming magnons with frequency $f_p/2$ and with opposite momenta are scattered into two outgoing magnons at the same frequency but with different momenta. These frequency-conserving scattering events only occur for parametrically injected magnons as indicated by light orange arrows in Fig. 2f; and *ii*) four-magnon scattering processes in which two magnons with a frequency of $f_p/2$ and different momenta scatter to two outgoing magnons with different frequencies and momenta. This frequency-nonconserving magnon scattering process involves all magnons in the system, i.e., both parametrically injected and scattered magnons, as indicated by red arrows in Fig. 2f.

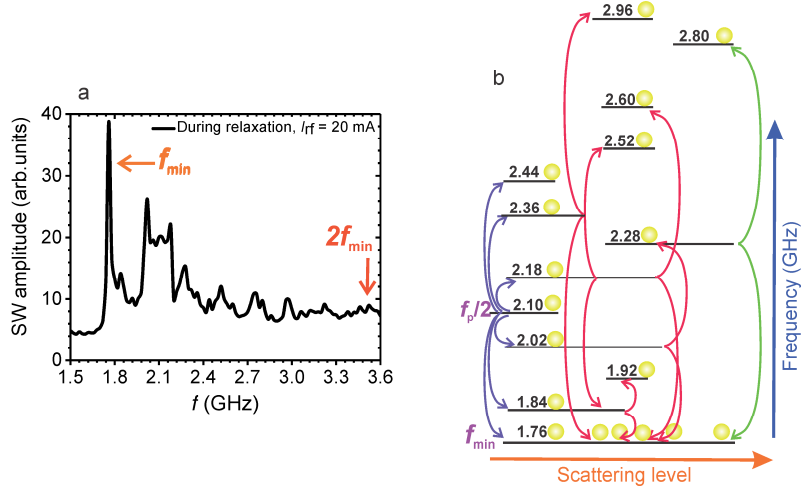

FIG. 3. (a) Frequency spectrum of the system during relaxation after it was pumped with microwave current of $I_{rf}$ = 20 mA. (b) Frequency levels involved in the four magnon scattering processes leading to the condensation at the band bottom. Numbers are frequencies of the magnons in GHz and correspond to the peaks displayed in the frequency spectrum shown in (a), blue, red and green arrows indicate individual frequency- nonconserving four magnon scattering processes at first, second and third levels, respectively. Each yellow circle represents a pair of magnons involved in the scattering process.

Such four-magnon scattering processes are essential for the thermalization of the magnon gas and formation of a magnon condensate at the bottom of the band [7-8]. Here, we should emphasize that we do not observe a high density of magnons at the band bottom during the pumping, see Fig. 2f. Furthermore, we note that since the lowest frequency $f_{min}$ in our system fulfills $2f_{min} > f_p/2$, three-magnon scattering events to the band bottom (where the condensation is expected) are prohibited.

In order to observe a notable amount of condensed magnon at the global energy minima $f_{min}$ at $k_x = \pm Q$, we increase the microwave pumping current further to $I_{rf}$ = 20 mA. We plot the band structure of the magnons during the pumping (Fig. 2g) and relaxation (Fig. 2h). Comparing Fig. 2g and Fig. 2h reveals how the energy redistribution occurs in the frequency-momentum space and how the parametrically injected magnons condense to the band bottom once the pumping is switched off.

Figure 2g demonstrates that the system under pumping possesses a strong redistribution of the energy to the entire spectrum, as discussed above. Once the pumping is switched off (Fig. 2h), most of the magnons start to condense at the two global minima of the magnon dispersion at $k_x = \pm Q$, as will be discussed in the following.

To elaborate on how this cascade of nonlinear four-magnon scattering events occur in the absence of pumping, we present the frequency spectrum of the conduit in Fig. 3a corresponding to Fig. 2h. The condensation at the band bottom is evident as a peak with the highest amplitude. Additionally, the peaks at higher frequencies indicate avalanche-like multi-level scattering events.

To further illustrate the details of these scattering mechanisms, in Fig. 3b we present the levels of the four-magnon scattering events in which the numbers correspond to the frequencies of magnon peaks as displayed in the frequency spectrum in Fig. 3a. Although higher-level scattering processes also occur, for simplicity, we only discuss the first, second and one of the third levels of scattering which are indicated by blue, red and green arrows, respectively.

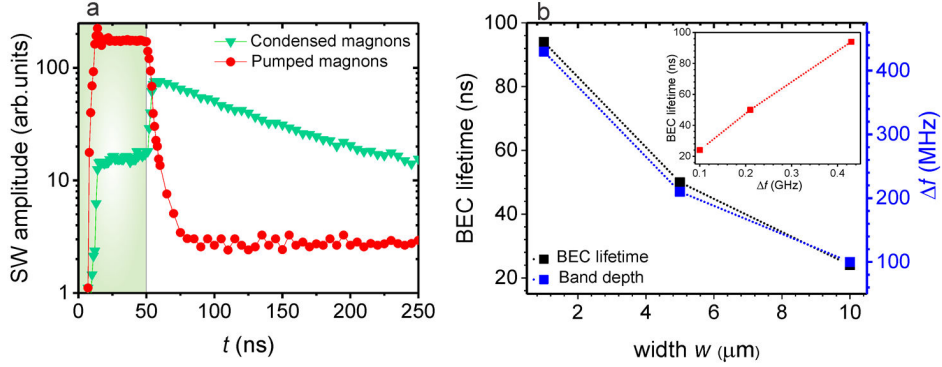

FIG. 4. (a) Amplitude of the parametrically generated magnons (red circles) and condensed magnons at the band bottom (green triangles) as a function of time. The pumping pulse is distinguished via the green background color. (b) The lifetime of the BEC (black color) and band depth $\Delta f$ with respect to the width of the conduits $w$. The inset shows the lifetime of the BEC with respect to the band depth $\Delta f$.

First, the parametrically generated magnons undergo a multi-level set of four-magnon scattering events (see blue arrows). Next, each scattered pair of magnons undergoes a second four-magnon scattering process individually as illustrated by the red arrows. This sequence continues to higher levels. For example, the third order as displayed by green arrows, shows scattering towards both higher frequencies and the band bottom. Finally, the population of the magnons at the band bottom increases significantly, and due to their long lifetime, a condensate at the band bottom forms.

In order to investigate the speed of magnon thermalization upon switching off the pumping, we present in Fig. 4a the temporal evolution of the amplitude of the magnons at $f_p/2$ (blue curve) and condensed magnons (red curve). We now let the system relax for $t_{\text{relax}} = 200$ ns. This figure demonstrates that during the pumping, the number of parametrically injected magnons increases rapidly, whereas the number of magnons at the band bottom increases slowly. Once the pumping is switched off, the parametrically injected magnons decay rapidly within only $t_{\text{decay}} = 5$ ns. However, in an opposite manner, the number of magnons at the band bottom increases dramatically within $t_{\text{rise}} \sim 5$ ns after the pumping has been switched off and only starts to decay afterwards. The thermalization time of approximately 5 ns that we observe is much faster than previous reports on bulk samples [8, 10]. This can be explained by the quantized spectrum of the system. In addition, the decay time of the condensed magnons at the band bottom is $t_{\text{decay}} = 94$ ns, which is shorter than the analytically calculated lifetime for linear (low amplitude) magnons at this spectral position ($t_{\text{lifetime}} \sim 220$ ns). We believe that such a difference is caused by the three-magnon confluence process of the condensed magnons as shown by $2f_{\text{min}}$ in Fig. 3a, in addition to nonlinear scattering processes near the band bottom as discussed above, which can open extra channels for magnon dissipation [8, 35].

It is important to mention that the condensate has a shorter lifetime and requires a longer thermalization time to form in a conduit with a larger width as shown in Fig. 4b and the supplemental materials [33]. This is caused by the fact that the laterally larger conduit has a shallower band depth (smaller $\Delta f$) and, consequently, a smaller energy distance between the condensed states at the bottom and the excited states [33], in agreement with our initial prediction. For example, as shown in the inset of Fig. 4b, the lifetime of the BEC in the $w = 10$ μm wide conduit with a $\Delta f = 0.1$ GHz is $t_{decay} = 24$ ns which is almost four times smaller than the BEC lifetime in the $w = 1$ μm conduit with $\Delta f = 0.43$ GHz.

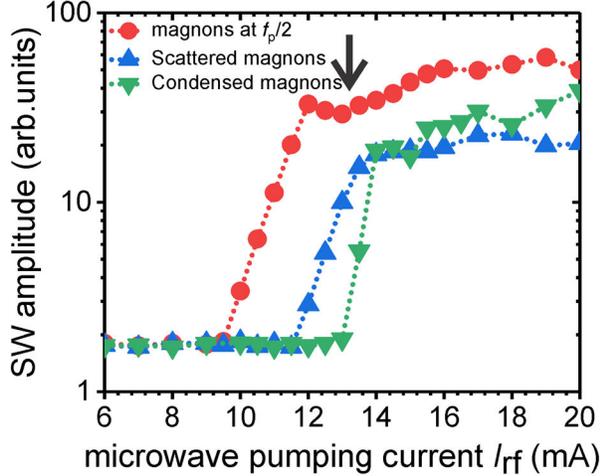

FIG. 5. Amplitude of the parametrically generated magnons (red circles), the scattered magnons via (frequency-nonconserving) four magnon scattering processes (blue triangles) both under pumping and condensed magnons under relaxation (green triangles) as a function of the microwave current.

We now vary the pumping field amplitude to study systematically the threshold dependence of the dynamics in the system. Figure 5 represents the amplitude of parametrically generated magnons during pumping (red circles), the nonlinear frequency-nonconserving four magnon scattering processes during the pumping (blue triangles), and the amplitude of condensed magnons during relaxation (green triangles). At $I_{rf} = 10$ mA, the amplitude of the generated magnons at $f_p/2$ starts to grow exponentially with the microwave field amplitude. This point is the onset of the parametric magnon generation instability [36, 37]. Increasing the microwave current increases the amplitude of the parametrically generated magnons. However, this trend changes at $I_{rf} = 11.5$ mA since a kink occurs in the growth rate of the parametrically generated magnons and their growth rate gets decreased by further increasing the current from this point, indicated by the black arrow [38]. This kink is caused by the onset of frequency-nonconserving four magnon scattering processes which start to grow at this microwave threshold current. More interestingly, the condensed magnon state also possesses a clear threshold, $I_{rf} = 13.5$ mA, which is above the threshold of the explained four-magnon scattering processes.

In conclusion, we have studied nonequilibrium magnon condensates in nanoscopic devices and proposed a new way to enhance their lifetime of by lateral confinement. We revealed the role of dipolar interactions in the generation of a magnon BEC as a metastable state in YIG ultrathin film structures. We showed that increasing the density of the parametrically injected magnons leads to the formation of a gaseous state involving weakly interacting magnons. Such a highly excited system establishes magnon condensation at the bottom of the lowest magnon mode (quantum ground state) during relaxation. Our results corroborate that the magnon condensation forms and its lifetime is enhanced if the two following criteria are met: *i*) the density of the magnons is high enough to allow the onset of frequency-nonconserving four-magnon scattering processes and formation a magnon gas, and *ii*) the band depth, i.e. the distance between the global minimum and the ferromagnetic resonance should be enhanced by lateral confinement. Our results determine that the thermalization time in a nanoscopic system with a diluted spectrum is much faster than in bulk samples with hybridized magnon mods. Our study provides a new way to obtain magnon condensation phenomena in nanoscopic confined geometries, and will motivate further experimental efforts to observe and use magnon condensates in integrated circuits for data transport and processing. Moreover, it will largely extend the freedom to investigate the dynamics behind magnon condensates, and further open an avenue to design suitable systems to exploit them in systems on a chip.

*Acknowledgement*--- This project is funded by the European Research Council within the Advanced Grant No. 694709 "SuperMagnonics", the Deutsche Forschungsgemeinschaft (DFG, German Research Foundation) - TRR 173 - 268565370 ("Spin+X", Project B01), and the Nachwuchsring of the TU

Kaiserslautern. A. Q. and A. B. are supported by the European Research Council via Advanced Grant No. 669442, "Insulatronics" and by the Research Council of Norway through its Centers of Excellence funding scheme, Project No. 262633, "QuSpin". Fruitful discussions with Q. Wang and H. Yu. Musiienko-Shmarova are appreciated.

*Correspondence to:*

*mohseni@rhrk.uni-kl.de*

**References**

[1] K. B. Davis, M. O. Mewes, M. R. Andrews, N. J. van Druten, D. S. Durfee, D. M. Kurn, and W. Ketterle, Bose-Einstein Condensation in a Gas of Sodium Atoms, Phys. Rev. Lett. **75**, 3969 (1995)

[2] M. H. Anderson, J. R. Ensher, M. R. Matthews, C. E. Wieman, E. A. Cornell, Observation of Bose-Einstein Condensation in a Dilute Atomic Vapor, Science **269**, 5221 (1995)

[3] L. V. Butov, A. L. Ivanov, A. Imamoglu, P. B. Littlewood, A. A. Shashkin, V. T. Dolgopolov, K. L. Campman, and A. C. Gossard, Stimulated scattering of indirect excitons in coupled quantum wells: signature of a degenerate Bose-gas of excitons. Phys. Rev. Lett. **86**, 5608–5611 (2001).

[4] J. Kasprzak, M. Richard, S. Kundermann, A. Baas, P. Jeambrun, J. M. J. Keeling, F. M. Marchetti, M. H. Szymańska, R. André, J. L. Staehli, V. Savona, P. B. Littlewood, B. Deveaud & Le Si Dang, Bose–Einstein condensation of exciton polaritons. Nature **443**, 409–414 (2006).

[5] J. Klaers, J. Schmitt, F. Vewinger & M. Weitz, Bose–Einstein condensation of photons in an optical microcavity. Nature **468**, 545–548 (2010).

[6] T. Giamarchi, C. Rüegg & O. Tchernyshyov, Bose–Einstein condensation in magnetic insulators, Nat. Phys. **4**, 198–204 (2008)

[7] S. O. Demokritov, V. E. Demidov, O. Dzyapko, G. A. Melkov, A. A. Serga, B. Hillebrands & A. N. Slavin, Bose–Einstein condensation of quasi-equilibrium magnons at room temperature under pumping. Nature **443**, 430–433 (2006)

[8] A. A. Serga, V. S. Tiberkevich, C. W. Sandweg, V. I. Vasyuchka, D. A. Bozhko, A. V. Chumak, T. Neumann, B. Obry, G. A. Melkov, A. N. Slavin & B. Hillebrands, Bose–Einstein condensation in an ultra-hot gas of pumped magnons, Nat. Comm. **5**, 3452 (2014)

[9] V. I. Yukalov, Difference in Bose-Einstein condensation of conserved and unconserved particles, Laser Physics **22**, 1145–1168 (2012)

[10] A.V. Chumak, V. I. Vasyuchka, A. A. Serga & B. Hillebrands. Magnon spintronics. Nat. Phys. **11**, 453–461 (2015)

[11] V. E. Demidov, O. Dzyapko, S. O. Demokritov, G. A. Melkov, and A. N. Slavin. Thermalization of a parametrically driven magnon gas leading to Bose–Einstein condensation. Phys. Rev. Lett. **99**, 037205 (2007).

[12] Chen Sun, Thomas Nattermann and Valery L Pokrovsky, J. Phys. D: Appl. Phys. **50** 143002 (2017).

[13] S. M. Rezende, Theory of coherence in Bose-Einstein condensation phenomena in a microwave-driven interacting magnon gas, Phys. Rev. B **79**, 174411 (2009)

[14] A. Rückriegel and P. Kopietz, Rayleigh-Jeans Condensation of Pumped Magnons in Thin-Film Ferromagnets, Phys. Rev. Lett. **115**, 157203 (2015)

[15] A. A. Serga, C. W. Sandweg, V. I. Vasyuchka, M. B. Jungfleisch, B. Hillebrands, A. Kreisel, P. Kopietz, and M. P. Kostylev, Brillouin light scattering spectroscopy of parametrically excited dipole-exchange magnons. Phys. Rev. B **86**, 134403 (2012).

[16] S. M. Rezende, Wave function of a microwave-driven Bose-Einstein magnon condensate, Phys. Rev. B **81**, 020414(R) (2010)

[17] T. Kloss, A. Kreisel, and P. Kopietz, Parametric pumping and kinetics of magnons in dipolar ferromagnets, Phys. Rev. B **81**, 104308 (2010)

[18] Q. Wang, B. Heinz, R. Verba, M. Kewenig, P. Pirro, M. Schneider, T. Meyer, B. Lägel, C. Dubs, T. Brächer, and A. V. Chumak. Spin Pinning and Spin-Wave Dispersion in Nanoscopic Ferromagnetic Waveguides. Phys. Rev. Lett. **122**, 247202 (2019)

[19] M. Mohseni, R. Verba, T. Brächer, Q. Wang, D. A. Bozhko, B. Hillebrands, and P. Pirro. Backscattering Immunity of Dipole-Exchange Magnetostatic Surface Spin Waves. Phys. Rev. Lett. **122**, 197201 (2019)

[20] D. A. Bozhko, A. A. Serga, P. Clausen, V. I. Vasyuchka, F. Heussner, G. A. Melkov, A. Pomyalov, V. S. L'vov, B. Hillebrands, Supercurrent in a room-temperature Bose-Einstein magnon condensate. Nat. Phys. **12**, 1027 (2016).

[21] D. A. Bozhko, A. J. E. Kreil, H. Y. Musiienko-Shmarova, A. A. Serga, A. Pomyalov, V. S. L'vov & B. Hillebrands, Bogoliubov waves and distant transport of magnon condensate at room temperature, Nat. Comm. **10**, 2460 (2019)

[22] B. Flebus, S. A. Bender, Y. Tserkovnyak, and R. A. Duine, Two-Fluid Theory for Spin Superfluidity in Magnetic Insulators, Phys. Rev. Lett. **116**, 117201 (2016)

[23] L. J. Cornelissen, K. J. H. Peters, G. E. W. Bauer, R. A. Duine, and B. J. van Wees. Magnon spin transport driven by the magnon chemical potential in a magnetic insulator. Phys. Rev. B **94**, 014412 (2016)

[24] A. Rückriegel and P. Kopietz, Spin currents, spin torques, and the concept of spin superfluidity, Phys. Rev. B **95** 104436 (2017)

[25] C. Ulloa, A. Tomadin, J. Shan, M. Polini, B. J. van Wees, and R. A. Duine. Nonlocal Spin Transport as a Probe of Viscous Magnon Fluids, Phys. Rev. Lett. **123**, 117203 (2019)

[26] A. Qaiumzadeh, H. Skarsvåg, C. Holmqvist, and A. Brataas, Spin Superfluidity in Biaxial Antiferromagnetic Insulators, Phys. Rev. Lett. **118**, 137201 (2017)

[27] M. Schneider, T. Brächer, V. Lauer, P. Pirro, D. A. Bozhko, A. A. Serga, H. Yu. Musiienko-Shmarova, B. Heinz, Q. Wang, T. Meyer, F. Heussner, S. Keller, E. Th. Papaioannou, B. Lägel, T. Löber, V. S. Tiberkevich, A. N. Slavin, C. Dubs, B. Hillebrands, and A.V. Chumak, Bose-Einstein condensation of quasi-particles by rapid cooling, arXiv:1612.07305 (2019), Nature Nano (2020)

[28] C. Safranski, I. Barsukov, H. K. Lee, T. Schneider, A. A. Jara, A. Smith, H. Chang, K. Lenz, J. Lindner, Y. Tserkovnyak, M. Wu & I. N. Krivorotov, Spin caloritronic nano-oscillator, Nature Commun. **8**, 117 (2017).

[29] B. A. Kalinikos, A. N. Slavin. Theory of dipole-exchange spin wave spectrum for ferromagnetic films with mixed exchange boundary conditions. Journal of Physics C: Solid State Physics **19** (35), 7013 (1986)

[30] I. S. Tupitsyn, P. C. E. Stamp, and A. L. Burin. Stability of Bose-Einstein Condensates of Hot Magnons in Yttrium Iron Garnet Films. Phys. Rev. Lett. **100**, 257202 (2008)

[31] F. L., Wayne M. Saslow & V. L. Pokrovsky, Phase Diagram for Magnon Condensate in Yttrium Iron Garnet Film. Sci. Rep. **3**, 1372 (2013)

[32] A. Vansteenkiste, J. Leliaert, M. Dvornik, M. Helsen, F. Garcia-Sanchez, and B. V. Waeyenberge, The design and verification of mumax3, AIP Advances **4**, 107133 (2014).

[33] See supplemental materials

[34] L'vov, V. S. Wave Turbulence under Parametric Excitation Springer (1994).

[35] P. Clausen, D. A. Bozhko, V. I. Vasyuchka, B. Hillebrands, G. A. Melkov, and A. A. Serga, Stimulated thermalization of a parametrically driven magnon gas as a prerequisite for Bose-Einstein magnon condensation, Phys. Rev. B **91**, 220402(R) (2015)

[36] M. Mohseni, M. Kewenig, R. Verba, Q. Wang, M. Schneider, B. Heinz, C. Dubs, A. A. Serga, B. Hillebrands, A. V. Chumak, P. Pirro, Parametric generation of propagating spin-waves in ultrathin yttrium iron garnet waveguides, Phys. Stat. Soli. RRL, **2000011,** (2020)

[37] T. Brächer, P. Pirro, B. Hillebrands. Parallel pumping for magnon spintronics: Amplification and manipulation of magnon spin currents on the micron-scale, Physics Reports **699**, (2017)

[38] P. Pirro, T. Sebastian, T. Brächer, A. A. Serga, T. Kubota, H. Naganuma, M. Oogane, Y. Ando, and B. Hillebrands, Non-Gilbert-damping Mechanism in a Ferromagnetic Heusler Compound Probed by Nonlinear Spin Dynamics, Phys. Rev. Lett. **113**, 227601 (2014)